\documentclass[12pt]{article}

\usepackage{amssymb,graphicx}

\usepackage{epsf}
\usepackage{graphicx,epsfig}
\usepackage{amsfonts}
\usepackage{amssymb}

\def\bk{{\bf k}}
\def\bp{{\bf p}}
\def\bx{{\bf x}}

\def\CA{{\cal A}}

\def\CO{{\cal O}}

\def\CG{{\cal G}}

\makeatletter
\renewcommand\section{\@startsection {section}{1}{\z@}%
                                 {-3.5ex \@plus -1ex \@minus -.2ex}
                                   {2.3ex \@plus.2ex}%
                                   {\normalfont\large\bfseries}}
\renewcommand\subsection{\@startsection{subsection}{2}{\z@}%
                                   {-3.25ex\@plus -1ex \@minus -.2ex}%
                                     {1.5ex \@plus .2ex}%
                                     {\normalfont\bfseries}}
\renewcommand\subsubsection{\@startsection{subsubsection}{3}{\z@}%
                                   {-3.25ex\@plus -1ex \@minus -.2ex}%
                                     {1.5ex \@plus .2ex}%
                                     {\normalfont\itshape}}
\makeatother

\newcommand{\Letter}{
\setlength{\textwidth}{16.5cm}
   \setlength{\textheight}{22.6cm}
    \hoffset=-0.5in
\voffset=-2.1cm }

\Letter

\begin{document}
\newcommand{\be}{\begin{equation}}
\newcommand{\ee}{\end{equation}}
\newcommand{\bea}{\begin{eqnarray}}
\newcommand{\eea}{\end{eqnarray}}
\newcommand{\barr}{\begin{array}}
\newcommand{\earr}{\end{array}}
\newcommand{\myfigure}[2]{\resizebox{#1}{!}{\includegraphics{#2}}}

\thispagestyle{empty}
\begin{flushright}
\parbox[t]{1.5in}{
astro-ph/yymmnnn}
\end{flushright}

\vspace*{0.3in}

\begin{center}
{\Large \bf Large Non-Gaussianities in Single Field Inflation}

\vspace*{0.5in} {Xingang Chen${}^1$, Richard Easther${}^2$ and Eugene A. Lim${}^2$ } 
\\[.3in]
${}^1$ Newman Laboratory,
Cornell University, Ithaca, NY 14853 \\
${}^2$   Department of Physics, Yale University, New Haven, CT 06511
 \\[0.3in]
\end{center}

\begin{center}
{\bf
Abstract}
\end{center}
\noindent
We compute the 3-point correlation function for  a general model of
inflation driven by a single, minimally coupled scalar field.  Our
approach is based on the  numerical evaluation of both the
perturbation equations and the integrals which contribute to the  3-point
function.  Consequently, we can analyze models where the potential has
a ``feature", in the vicinity of which the slow roll parameters may
take on large, transient values.  This introduces both scale and shape
dependent non-Gaussianities into the primordial perturbations.   As an
example of our methodology,  we examine the ``step'' potentials which
have been invoked to improve the fit to the glitch in the $\langle TT
\rangle$ $C_\ell$ for $\ell \sim 30$, present in both the one and
three year WMAP data sets. We show that for the typical parameter
values, the non-Gaussianities associated with the step are far larger
than those in standard slow roll inflation, and may even be within reach of
a next generation CMB experiment such as Planck.  More generally, we
use this example to explain that while adding features to potential
can improve the fit to the 2-point function, these are {\em
generically\/} associated with a greatly enhanced signal at the
3-point level. Moreover, this 3-point signal will have a very
nontrivial shape and scale dependence, which is correlated with the
form of the 2-point function, and may thus lead to a consistency check on models of inflation with non-smooth
potentials. 

\vfill

\newpage
\setcounter{page}{1}


\newpage

\section{Introduction}

The Cosmic Microwave Background is a treasure trove of information
about the primordial universe. An exciting set of current and future
experiments \cite{Spergel:2006hy,Scoccimarro:2003wn,Liguori:2005rj} 
will yield an even richer data set, allowing us to
make precise tests of inflationary models via their predictions for
the  form of the primordial perturbations.   Typically, these
predictions are expressed in terms of the power spectrum, or 2-point
function. For a general set of initial inhomogeneities, further information about the perturbations might be
gleaned from their higher order correlation functions.   If the perturbations are exactly Gaussian, then the N-point functions vanish exactly when $N$ is odd, and is fully specified in terms of the 2-point function for even $N$.

All inflationary models predict {\em some\/} level of non-Gaussianity. Moreover,
non-linear corrections will generate non-Gaussianities in the CMB even
if the primordial spectrum is purely Gaussian \cite{Pyne:1995bs}. Thus
any observed non-Gaussianities will be a combination of the
primordial non-Gaussianity laid down during inflation, and
that arising from weak second-order couplings between modes. A rough measure of the non-Gaussianities is provided by the parameter
$f_{NL}$,
\begin{equation}
\Phi(x)=\Phi_L(x) + f_{NL}(\Phi_L(x)^2-\langle \Phi_L(x)^2\rangle)
\label{eqn:fnlansatz}
\end{equation}
where $\Phi(x)$ is the gravitational potential which sources the
temperature anisotropies in the CMB
\cite{Komatsu:2001rj}. Generically, second order couplings between
modes yield $f_{NL}\approx O(1)$,  while it has been shown in great
detail that for standard single field slow roll inflation, $f_{NL} \sim
\epsilon \sim O(10^{-2})$, where $\epsilon$ is the usual slow roll
parameter \cite{Maldacena:2002vr}.   The two contributions are cumulative,  and the primordial 3-point function is thus swamped by the
evolutionary contribution. Moreover, cosmic variance ensures that even
a perfect CMB dataset will not allow a conclusive detection of the
3-point function if $f_{NL}$ is of order unity.   In addition, the
presence of foreground contamination \cite{Scoccimarro:2003wn,Verde:1999ij} further complicates
observational efforts. Current estimates suggest that Planck  can
achieve a limit of $|f_{NL}|<20\sim30$ \cite{Hikage:2006fe}. In this paper paper we will show that
this may permit the detection of
interesting features in the 3-point statistics for some classes of
scale-dependent inflationary models.

The 3-point correlation function is a separate and
independent statistic. Therefore
it potentially discriminates between models of
inflation with degenerate power spectra. More optimistically, if the amplitude of the
3-point function is large enough for us to \emph{map} its dependence
on both the \emph{scale} and \emph{shape} of the momenta triangle,
this will be an enormous boon to early universe cosmology,
since the 3-point function  encodes information in both these properties.   
However, to  realize this
possibility, we need to compute the 3-point
correlation function for a given inflationary model, and compare  these
predictions to data.  Here we show how to compute the
3-point function for a  general model of inflation driven by a single,
minimally coupled scalar field. Crucially, we do not assume slow roll,
and our methodology applies to both simple models with smooth
potentials, and  more complicated scenarios where the potential
has an isolated ``feature''.   Since the 3-point function can have
both shape and scale dependence, our calculations will underscore the
limitations of a single scalar statistic such as $f_{NL}$.  Our
analysis is purely theoretical, and we do not   compare our computed 3-point
function to data, but this issue is being actively addressed by others
\cite{Kendrick}. 

As
cosmological data improves, we will frequently be faced with the
problem of interpreting ``glitches'' in the observed power spectra,
and determining whether these represent genuine departures from some
concordance cosmology.  For instance, the $C_\ell$ derived from the
temperature anisotropies in both the 1 Year and 3 Year WMAP datasets
show a noticeable departure from the best fit LCDM spectra around
$\ell \sim 30$, and the fit can be improved by adding a small ``step''
to the inflationary potential.  The best-fit parameters for this step
were extracted from the 1 Year data set in \cite{Peiris:2003ff}, and
from the 3 Year dataset in \cite{Covi:2006ci}, and the two analyses
find broadly similar values.  Of course, any proposal that adds free
parameters to the inflationary potential is likely to produce a better
fit to data. The question then turns to whether this improvement is
good enough to justify the additional parameters. If we focus solely
on the 2-point function, this question is addressed via the
resulting change in the $\chi^2$ per degree of freedom, or by Bayesian
evidence.  However, we will see that a potential with a sharp,
localized feature induces a massive enhancement of the primordial
3-point function, relative to our expectations from slow-roll. This enhancement is sufficiently dramatic to boost the
non-Gaussianities to the point where they may be detected experimentally. Consequently, this provides a
stringent check on any proposal for a nontrivial
inflationary potential, since the specific form of the
non-Gaussianities will be heavily correlated with the 2-point
function. Such correlations will also extend to higher order statistics such as the trispectrum and beyond, though we do not consider them here \cite{Okamoto:2002ik,Kogo:2006kh,Byrnes:2006vq,Seery:2006js,Seery:2006vu,Hu:2001fa,Huang:2006eh}.

In models with a non-trivial potentials, an accurate computation of the power spectrum must be performed numerically. We begin  our examination of the 3-point function by
reproducing existing semi-analytical results that
assume slow roll.  We then analyze a
specific model where the inflaton potential has a small, sharp
``step'' \cite{Adams:2001vc} to illustrate our methodology in a
non-trivial setting.  Since this model has already been
used to improve the fit between LCDM cosmology and the observed power
spectrum, we can take the best-fit parameter values for this
step and compute the resulting 3-point function.
 Interestingly, although such a feature only causes a small correction in the
2-point correlation function , it amplifies the non-Gaussianity by a factor around 1000. Roughly speaking\footnote{As we will discuss later, $f_{NL}$ is not a good statistic for such non-trivial non-Gaussianities, although it remains useful as a rough guide to the amplitude of the 3-point function.},  $f_{NL}$ is boosted from $\CO(.01)$ to $\CO(10)$,  making the non-Gaussianity a potentially useful constraint.  The leading term in the cubic Lagrangian responsible for this boost differs from those used in the previous calculations
in Ref.~\cite{Maldacena:2002vr,Seery:2005wm,Chen:2006nt}.

The distinctive feature of this non-Gaussianity is its
dependence on the scale of the momenta triangle
\cite{Chen:2005fe}. 
Firstly, it has a characteristic ``ringing'' behavior; as we will see, a suitably defined generalization of $f_{NL}$ oscillates between positive
and negative values. Secondly, the impact on the 3-point function is
localized around the wavenumbers that are most affected by any
feature, and dies away over several 
e-folds. These features distinguish a step model from other single field
theories with large non-Gaussianities, such as DBI
inflation \cite{Silverstein:2003hf,Chen:2004gc} or k-inflation models
\cite{Armendariz-Picon:1999rj}, where the non-Gaussianities are
present at all scales and the running is very slow
\cite{Alishahiha:2004eh,Chen:2005fe,Chen:2006nt,Shandera:2006ax,Kecskemeti:2006cg}.
 
This paper is divided into the following sections. Section
\ref{section:model} details the step potential model that we are
considering; we lay out our conventions in this section. Section
\ref{section:3pts} details the analytic forms of the 3-pt correlation
functions, which we will proceed to integrate numerically in Section
\ref{section:numerics}. We discuss our results in Section
\ref{section:results} and conclude in Section
\ref{section:conclusion}.

\section{Slow roll model with a feature} \label{section:model}

For single field inflation driven by a minimally coupled scalar, the action is  
\begin{equation}
S = \int dx^4 \sqrt{g} \left[\frac{M_p^2}{2} R + \frac{1}{2}(\partial \phi)^2 - V(\phi)\right] \label{eqn:backgroundaction}
\end{equation}
where the potential $V(\phi)$ is designed such that the inflaton field
$\phi$ is slowly rolling for long enough to drive inflation. We define the slow
roll parameters by\footnote{Note these are related to the potential slow roll parameters via $\epsilon_V
= (M_p^2/2) (V'/V)^2$, $\eta_V = M_{p}^2
V''/V$, by $\epsilon=\epsilon_V$, $\eta=-2\eta_V + 4\epsilon_V$.}
\begin{eqnarray}
\epsilon&=&\frac{\dot{\phi}^2}{2M_p^2 H^2} \label{eqn:epsilon} ~ ,\\
\eta &=& \frac{\dot{\epsilon}}{\epsilon H} \label{eqn:eta} \, .
\end{eqnarray}
For the quadratic potential,
$V(\phi) = m^2\phi^2 /2$, $\eta = 2\epsilon$. Assuming
transplanckian field values, or $\phi>M_p$, both slow roll parameters
and their higher derivative cousins are sub-unity.

We add a step into the slow roll potential, with the form proposed by \cite{Adams:2001vc}
\begin{equation}
V(\phi)=\frac{1}{2}m^2\phi^2\left[1+c \tanh\left(\frac{\phi-\phi_s}{d}\right)\right], \label{eqn:potential}
\end{equation}
where the step is at $\phi_s$, with size $c$ and gradient
$d$ respectively. While the presence of the step at just the ``right''
place would require a tuning, one can imagine potentials with many
steps, so that the odds are high that at least one of them would fall
inside the range of $\phi$ relevant to the cosmological perturbations.

\begin{figure}[t]
\myfigure{3in}{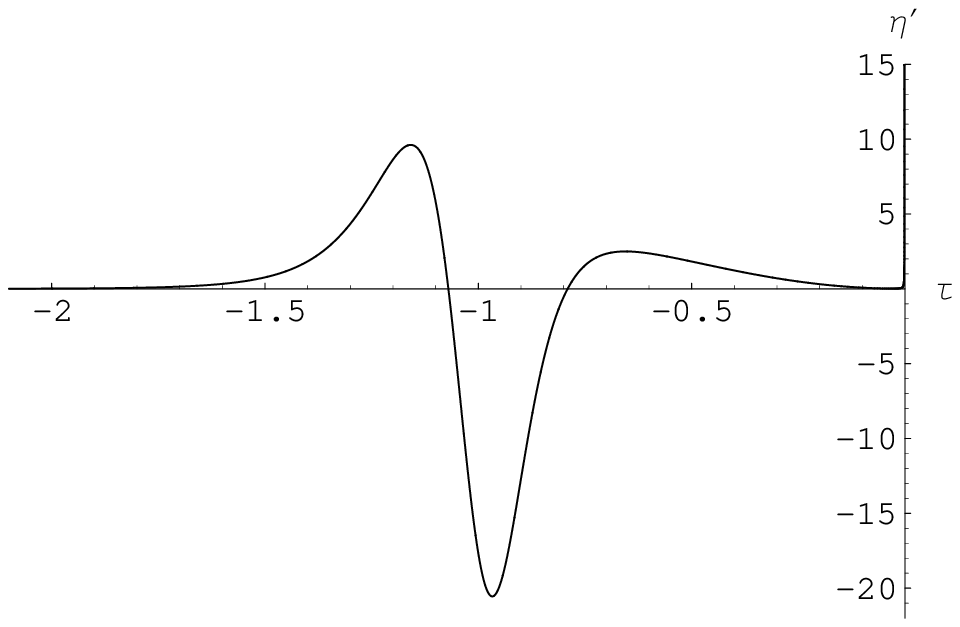}
\myfigure{3in}{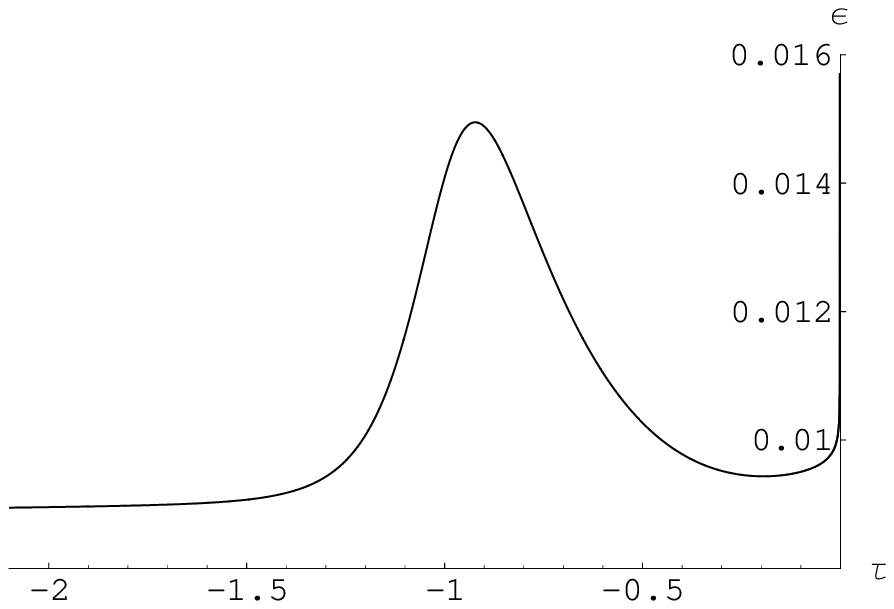}
\caption{The $\eta'$ (left) and $\epsilon$ (right) evolution over the
step for the model $c=0.0018,~d=0.022$, with its amplitude
momentarily $\eta'\approx 10c^2/(\epsilon d^2)$. This is the
primary source of the large non-Gaussianities in the step potential, as the leading term in the 3 point expansion is of order $\eta' \epsilon$. On the other hand, since the height of the step is small, so the ratio of the kinetic energy to the potential energy
remains tiny and $\epsilon \ll 1$.}  
\label{fig:etadotplot}
\end{figure}

As discussed in \cite{Starobinsky:1992ts,Adams:2001vc} this step induces an
oscillatory ringing in the power spectrum. When the inflaton rolls through the step, it undergoes a strong momentary acceleration. In realistic models the step is typically less than
1\% of the overall height of the potential.   In this case,
$\dot{\phi}^2 \ll V(\phi)$ at all times, and $\epsilon \ll 1$.  On the
other hand, $\eta \propto V''$ and its higher derivatives can grow
dramatically as the inflaton crosses the step, often to the point
where they exceed unity, as shown in Fig.~(\ref{fig:etadotplot}). 

We begin with an order of magnitude estimate of
the values of the slow roll parameters. 
The step  has a depth
$\Delta V \approx \tanh (1) c  m^2 \phi^2$ and a width $\Delta \phi
\approx 2d$ (setting $M_{p}=1$). When the inflaton falls down the
step,   $\Delta V$  of the potential energy is converted to kinetic energy,
resulting in an increase in $\epsilon$ on the order of $\Delta \epsilon \approx \Delta
V/H^2 \approx 5c$, within a time interval $\Delta t \approx \Delta \phi/
\dot \phi \approx d/\sqrt{cV}$. Using these quantities, we can
estimate
\begin{equation}
\eta = \frac{\dot \epsilon}{H \epsilon} 
\approx \frac{7 c^{3/2}}{d\epsilon}
\end{equation}
and
\bea
\dot \eta &=& H (\frac{\ddot \epsilon}{H^2\epsilon} + \epsilon \eta -
\eta^2) \nonumber \\
&\approx& H \frac{10c^2}{d^2 \epsilon} 
\left( 1+ \frac{0.7d\epsilon}{\sqrt{c}} - \frac{4.5c}{\epsilon} \right)
\nonumber \\
&\approx& H \frac{10c^2}{d^2 \epsilon} ~.
\eea
These values have the same order of magnitude as the peaks obtained 
numerically in
Fig.~\ref{fig:etadotplot}.

After this acceleration, the inflaton is damped by the friction term in
\bea
\ddot \phi + 3H\dot \phi + \frac{dV}{d\phi} =0 
\label{eqn:attractor}
\eea
and relaxes to the attractor solution. This relaxation time is
determined by the first two terms in Eq.~(\ref{eqn:attractor}), 
and is of
order  $H^{-1}$. This corresponds to
the {\em decay} 
width of the peaks in Fig.~\ref{fig:etadotplot}.  It is easy to see that
$\eta$ and $\dot \eta$ during the relaxation are of order $\CO(1)$ and
$\CO(H)$, respectively. Our numerical
results were  obtained with $c=0.0018$, $d=0.022$ -- the central values
of the step corresponding to the low-$\ell$ glitch analyzed by Covi
{\em et al.\/}   \cite{Covi:2006ci}. Thus $\epsilon \approx
2/\phi^2\approx 2/14.8^2$, and get $\eta' \approx
10c^2/(d^2\epsilon) \approx 7$,
consistent with Fig.~\ref{fig:etadotplot}.
 Note that $aH\approx -1/\tau$ is chosen
to be around 1 in the plot, $\tau$ is the conformal time, and prime is
the derivative with respect to $\tau$.
 
The non-linear couplings of the perturbations are proportional
to powers of the slow roll parameters which characterize the
background evolution, even when the slow roll parameters are large. Consequently, a
deviation from slow roll results in a large mixing (``interaction'')
of modes, and large non-Gaussianities. 
The amplification of the non-Gaussianity relative to a smooth potential is
roughly characterized by $\dot \eta_{feature} \Delta
t_{feature}/\epsilon$. From our previous discussion we know that the
$\dot \eta \Delta t$ during relaxation after traversing the feature is of order
$1$. During acceleration this quantity will depend on the firm of the feature, and is greater than unity for  the step potential.  Thus any sharp feature will  greatly boost the 
non-Gaussianity.

\section{3-point correlation functions} \label{section:3pts}

In this section, we sketch out the derivation of the 3-point
correlation function, following
\cite{Maldacena:2002vr,Seery:2005wm,Chen:2006nt}, writing in terms of integrals over conformal time $\tau$, with the
integrand being a finite expansion in slow roll parameters. 
Although we work with the step potential, the analysis is applicable to 
a generic feature. We begin by perturbing the fields in the ADM metric \cite{Arnowitt:1960es}
\bea
ds^2 = -N^2dt^2 + h_{ij} (dx^i + N^i dt) (dx^j + N^j dt)
\eea
using the comoving gauge \cite{Mukhanov:1990me,Maldacena:2002vr}
\bea
h_{ij} = a^2 e^{2\zeta} \delta_{ij} ~,
\eea
where $N$ and $N^i$ are Lagrangian multipliers, and $\zeta$ is the
scalar perturbation. Note that in this gauge $\delta \phi$ vanishes. The power spectrum is given by the 2-point correlation function of the
curvature perturbation 
\begin{eqnarray}
\langle \zeta(\bx)\zeta(\bx)\rangle &=&\int \frac{d^3k_1}{(2\pi)^3}
\frac{d^3k_2}{(2\pi)^3}\langle \zeta(\bk_1) \zeta(\bk_2)\rangle
e^{i(\bk_1+\bk_2)\cdot \bx} \nonumber \\
&=& \int \frac{dk}{k}P_{\zeta} ~,
\end{eqnarray}
where $P_{\zeta}$ is the power spectrum.

The 3-point correlation function is substantially more complicated. As non-Gaussianities arise from
departures from nonlinear couplings, we must
compute the action (\ref{eqn:backgroundaction}) to cubic order in the
perturbation: 
\begin{eqnarray}
S_3&=&\int dtd^3x\{
a^3\epsilon^2 \zeta\dot{\zeta}^2+a\epsilon^2 \zeta(\partial\zeta)^2-
2a \epsilon\dot{\zeta}(\partial
\zeta)(\partial \chi) \nonumber \\ && +
\frac{a^3\epsilon}{2}\frac{d\eta}{dt}\zeta^2\dot{\zeta}
+\frac{\epsilon}{2a}(\partial\zeta)(\partial
\chi) \partial^2 \chi +\frac{\epsilon}{4a}(\partial^2\zeta)(\partial
\chi)^2+ 2 f(\zeta)\frac{\delta L}{\delta \zeta}|_1 \} ~,
\label{cubicaction}
\end{eqnarray}
where
\begin{eqnarray}
\chi &=& a^2 \epsilon \partial^{-2} \dot \zeta ~, \\
\frac{\delta
L}{\delta\zeta}\mid_1 &=& a
\left( \frac{d\partial^2\chi}{dt}+H\partial^2\chi
-\epsilon\partial^2\zeta \right) ~, \\
f(\zeta)&=&\frac{\eta}{4}\zeta^2+ {\rm terms~with~derivatives~on~\zeta
}
~.
\label{redefinition}
\end{eqnarray}
Here $\partial^{-2}$ is the inverse Laplacian and $\delta
L/\delta\zeta|_1$ is the variation of the quadratic action with
respect to the perturbation $\zeta$.  The cubic action
(\ref{cubicaction}) is exact for arbitrary $\epsilon$ and $\eta$.  Note that
the highest power of slow-roll parameters which appears is of
$\CO(\epsilon^3)$, when we recall that $\chi$ contains a
multiplicative factor of $\epsilon$. The fact that the series
expansion in
slow-roll parameters is finite is important since, although $\epsilon$
remains small in the presence of a feature, $\eta$ and $\eta'$ may become large.
The last term in  (\ref{cubicaction})  can be absorbed by a field redefinition of
$\zeta$,
\begin{eqnarray}
\zeta \rightarrow \zeta_n+f(\zeta_n) ~. \label{eqn:fieldredef}
\end{eqnarray}
After this redefinition, the interaction Hamiltonian in conformal time is 
\begin{eqnarray}
H_{int}(\tau) &=&
-\int d^3x \Bigg\{ a \epsilon^2 \zeta \zeta'^2  + a\epsilon^2 \zeta (\partial \zeta)^2  - 2 \epsilon \zeta' (\partial \zeta) (\partial \chi) 
\nonumber \\
&& + \frac{a}{2} \epsilon \eta' \zeta^2 \zeta'
+ \frac{\epsilon}{2a} (\partial \zeta) (\partial \chi) 
(\partial^2 \chi)
+ \frac{\epsilon}{4a} (\partial^2 \zeta) (\partial \chi)^2
\Bigg\} ~.
\label{Hint3}
\end{eqnarray}
It is more convenient to work in Fourier space, so
\begin{equation}
\langle \zeta(\bx)\zeta(\bx)\zeta(\bx)\rangle = \int \frac{d^3k_1}{(2\pi)^3} \frac{d^3k_2}{(2\pi)^3} \frac{d^3k_3}{(2\pi)^3} \langle \zeta(\bk_1)\zeta(\bk_2)\zeta(\bk_3)\rangle e^{i(\bk_1+\bk_2+\bk_3)\cdot \bx}.
\end{equation}
The field redefinition equation (\ref{eqn:fieldredef}) introduces some extra terms in the 3-point correlation function,
\begin{eqnarray}
\langle\zeta(\bk_1)\zeta(\bk_2)\zeta(\bk_3)\rangle
&=&\langle\zeta_n(\bk_1)\zeta_n(\bk_2)\zeta_n(\bk_3)\rangle
\nonumber \\
&+&\eta
\langle \zeta_n^2(\bk_1) \zeta_n(\bk_2) \zeta_n(\bk_3) \rangle
+ {\rm sym} + \CO(\eta^2 (P^\zeta_k)^3) ~,
\label{3ptredef}
\end{eqnarray}
where $\zeta_n^2(\bk)$ denotes the Fourier transform of
$\zeta_n^2(\bx)$ -- note that this is not the square of
$\zeta_n(\bk)$. In (\ref{3ptredef}), 
the slow roll parameters are evaluated at the end of inflation.
Although we can no longer perturbatively expand in terms of $\eta$ if
$\eta$ is large, the terms that we neglected are of higher order in
$P^\zeta_k$. Since $P^\zeta_k \sim 10^{-10}$, we can neglect these
unless  $\eta$ becomes ridiculously large, which it never does in any case of interest to us.  We have only
considered the first term in (\ref{redefinition}), since all other
terms involve at least one derivative of $\zeta$, and vanish when
evaluated outside the horizon. 

The 3-point correlation function at some time $\tau$ after horizon exit is then the vacuum expectation value of the three point function in the interaction vacuum
\begin{equation}
\langle\zeta(\tau,\textbf{k}_1)\zeta(\tau,\textbf{k}_2)\zeta(\tau,\textbf{k}_3)\rangle=
-i\int_{\tau_0}^{\tau} d\tau' ~ a ~ \langle
[
\zeta(\tau,\textbf{k}_1)\zeta(\tau,\textbf{k}_2)\zeta(\tau,\textbf{k}_3),{H}_{int}(\tau')]
\rangle ~. \label{eqn:interaction}
\end{equation}
Before we proceed, let us pause and point out a subtlety hidden inside
equation (\ref{eqn:interaction}): the 3-point on the left hand side is
evaluated with the \emph{interaction} vacuum while the the right hand
side is evaluated at the ``true'' vacuum.
The interaction Hamiltonian
${H}_{int}$ evolves the true vacuum to the interaction vacuum at the
time we evaluate the 3-point function. Neglecting this will lead to
errors as first pointed out by Maldacena
\cite{Maldacena:2002vr,Gangui:1993tt,Acquaviva:2002ud}.

The three terms in the second line of (\ref{Hint3}) 
are of higher order in slow-roll parameters, and were properly neglected
in Refs.~\cite{Maldacena:2002vr,Seery:2005wm,Chen:2006nt}, which  assume that $\eta$ and $\epsilon$ are always small. In our calculation, 
the $\epsilon^3$ terms are small, but the $\epsilon
\eta'$ term may be large and in fact dominant. 
the step.  

We now decompose $\zeta$ into its Fourier modes and quantize it by writing
\begin{equation}
\zeta(\tau,\bx) = \int \frac{d^3 \bp}{(2\pi)^3} \zeta(\tau,\bk) 
e^{i \bp \cdot \bx} ~,
\end{equation}
with associated operators and mode functions
\begin{eqnarray}
\zeta(\tau,\textbf{k})=u(\tau,\textbf{k})a(\textbf{k})
+u^*(\tau,-\textbf{k})a^{\dagger}(-\textbf{k}) ~,
\end{eqnarray}
where $a$ and $a^\dagger$ satisfy the commutation relation
$[a(\bk),a^\dagger(\bk')]=(2\pi)^3 \delta^3(\bk-\bk')$. The ``true''
vacuum is annihilated by the lowering operator $a(\bk)$. 
The power spectrum is then
\begin{equation}
P_{\zeta} \equiv \frac{k^3}{2\pi^2} |u_\bk|^2 ~.
\end{equation}
with $u_\bk$ is $u(\tau, \bk)$ evaluated after each mode crosses the
horizon. 

Meanwhile $v(\tau,\bk)$ is the solution of the linear equation of motion of the quadratic action,
\begin{equation}
v_k'' + k^2 v_k - \frac{z''}{z} v_k =0 ~, \label{eqn:mukhanov}
\label{quadeom}
\end{equation}
where we have used the definitions
\begin{equation}
v_k\equiv z u_k ~, ~~~~z \equiv a \sqrt{2\epsilon} ~.
\label{vdef}
\end{equation}
Our choice of vacuum implies that the initial condition for the mode function is given by the Bunch-Davies vacuum
\begin{eqnarray}
v_k(\tau_0)&=& \sqrt{\frac{1}{2k}}  \nonumber \\
v_k'(\tau_0)&=& -i\sqrt{\frac{k}{2}} \label{eqn:initialconditions}
\end{eqnarray}
where we have neglected an irrelevant phase, since  (\ref{quadeom}) is rotationally invariant. 

To compute the 3-point correlation function, one simply substitutes
these solutions into equation (\ref{Hint3}), and integrates the mode
functions from $\tau_0$ through to the end of inflation. This integral
can be done semi-analytically 
for simple models, provided the slow roll parameters are
small and relatively constant. Any departures from this serene picture
will render the semi-analytic approach intractable -- including the step potential we consider here. 

Inspecting equations (\ref{Hint3}) and (\ref{eqn:interaction}) and
recalling that $\chi_k \propto \epsilon \dot \zeta_k$, it is clear that the 3-point correlation function consists of a sum of integrals of the form  
\begin{equation}
I_{\epsilon^2} \propto \Re \left[\prod_i
u_i(\tau_{end})\int_{\tau_0}^{\tau_{end}}d\tau \epsilon^2 a^2
\xi_1(\tau)\xi_2(\tau)\xi_3(\tau) + \CO(\epsilon^3) \right] \label{eqn:Itype1}
\end{equation}
\begin{equation}
I_{\epsilon\eta'} \propto  \Re \left[\prod_i
u_i(\tau_{end})\int_{\tau_0}^{\tau_{end}}d\tau \epsilon\eta' a^2
\xi_1(\tau)\xi_2(\tau)\xi_3(\tau) \right] \label{eqn:Itype2} 
\end{equation}
where $\xi_n$ is either $u_{k_n}^*$ or $du_{k_n}^*/d\tau$. In a single
field model $u_{k}(\tau)\rightarrow \textrm{const}$ after Hubble
crossing as it freezes out, while $u_{k}(\tau)\rightarrow e^{-ik\tau}$
oscillates rapidly at early times, so its contribution to the integral tends to cancel. Thus the integral is dominated by the range of $\tau$ during which the modes leave the horizon. 

Now let us consider the potential (\ref{eqn:potential}). We expect
$\epsilon \ll 1$,  making (\ref{eqn:Itype1}) negligible.
However, $\eta$ and $\eta'$ can clearly be very large
as $H$ changes rapidly over a very short time. This flagrant
violation of slow roll means that we expect a large contribution from
the $I_{\epsilon\eta'}$ term in the integral, of the order
$\Delta \tau \eta'/\epsilon \times I_{\epsilon^2}$, where $\Delta \tau$ is
the acceleration time for the inflaton by the feature. From the qualitative estimation
in Sec.~\ref{section:model}, we know this is $7c^{3/2}/(d\epsilon^2)
\times I_{\epsilon^2}$.
The best parameters from Covi {\em et al.\/} for a step that matches
the low-$\ell$ glitch seen in current CMB data are  $c=0.0018$ and
$d=0.022$  \cite{Covi:2006ci}. Thus we expect a boost in the 3-point
statistic of $\CO(100\sim1000)$ at scales affected by the
step. Given that the baseline  3-point function is
proportional to $\epsilon$, we see that this pushes the expected value
up to $\CO(10)$ from $\CO(10^{-2})$.

We now focus on the $I_{\epsilon\eta'}$ term, and relegate the
discussion of the  subleading terms to an appendix
(where a field
redefinition term is also subleading since the effect of the feature
dies away towards the boundary). So our leading term is
\begin{equation}
i \left( \prod_i u_{i}(\tau_{end}) \right)  \int_{-\infty}^{\tau_{end}} d\tau a^2 \epsilon  \eta' 
\left( u_{1}^*(\tau) u_{2}^*(\tau) \frac{d}{d\tau} u_{3}^*(\tau)
+ {\rm two~perm} \right) (2\pi)^3 \delta^3(\sum_i \bk_i) + {\rm c.c.}
~,
\nonumber\\
\label{term1}
\end{equation}
where the ``two perm'' stands for two other terms that are symmetric under
permutations of the indices 1, 2 and 3, where 1, 2, 3  are shorthand for $k_1$,
$k_2$ and $k_3$.

We  must numerically solve the equations of
motion to get $u_k(\tau)$, and then evaluate
(\ref{term1}). The contributions from the other five terms in the
interaction Hamiltonian (\ref{Hint3}) in the appendix can be
calculated in a similar fashion. 
 
\section{Numerical Method} \label{section:numerics}

  The equations of motion for the background
fields and the perturbations are
\begin{eqnarray}
H(\tau)& =& \frac{a'}{a^2}, \\
\frac{d^2 a}{d\tau^2} &=& \frac{ a}{6 M_p^2}\left(-\phi'^2+4a^2V(\phi)\right),\\
\frac{d^2 \phi}{d\tau^2}&=&-2\frac{a'}{a}\phi'-a^2\frac{dV}{d\phi},  \\
\frac{d^2 v_k}{d\tau^2}& = & -\left(k^2-\frac{z''}{z}\right)v_k, 
\end{eqnarray}
where  $\tau$ is our time parameter. We choose initial conditions and
units for $\tau$ such that the step occurs around $\tau=-1$, and the
mode with comoving wavenumber $k=1$ crosses the horizon at the same point. The initial
conditions of the perturbations are given by
(\ref{eqn:initialconditions}).  

We need to evaluate  (\ref{term1}) and equations (\ref{term2}) to
(\ref{term4}). They all have the form (\ref{eqn:Itype1}) and
(\ref{eqn:Itype2}).   At  early times the integrand is highly
oscillatory and the net contribution per period is thus very small.  However the early time limit of the integral must be handled
with care, since a sharp cutoff imposed by a finite initial time will
introduce a spurious contribution of $\CO{(1)}$ into the result.  

Analytically, the standard procedure is to Wick rotate the integral
slightly into the imaginary plane
\cite{Maldacena:2002vr,Seery:2005wm,Chen:2006nt} in order to eliminate
the oscillatory terms. Unfortunately, it is not straightforward to
implement this approach numerically.  Instead we introduce a ``damping'' factor $\beta$ into the integrand 
\begin{equation}
I_{\epsilon^2} \propto \int_{\tau_{early}}^{\tau_{end}} d\tau a^2 \epsilon^2 \xi_1\xi_2\xi_3 \times e^{\beta(k_1+k_2+k_3) (\tau-\tau_{end}) },
\end{equation}
and similarly for the $I_{\eta'\epsilon}$ term.
This is approximately equivalent to rotating the contour integral
from the real axis slightly into the imaginary plane $\tau \rightarrow
\tau(1-i\beta)$. 
The relative error introduced by this numerical procedure can be estimated by
taking the difference between the integration over $e^{i(k_1+k_2+k_3)\tau}$ and
$e^{(i+\beta)(k_1+k_2+k_3)\tau}$ and is thus $\beta$.
If we ensure
\begin{eqnarray}
|\beta k \tau_{crossing}|& \ll & 1 ~,\\
|\beta k \tau_{early}| &\gg & 1 ~,
\end{eqnarray}
then this contribution is negligible, recalling that $\tau<0$.  

We can adjust both $\tau_{early}$ and $\beta$.  Choosing an appropriate combination will  damp out the oscillatory
contributions at early times while  having no effect on the integrand during
Hubble crossing, which provides the dominant contribution to the
3-point function.  We tested this scheme against the known analytical
results for 3-point functions, which we match to within a few percent.    However,  $\beta$ and $ \tau_{early}$ are set by hand in our code, so while this approach is suitable for an
initial survey, we are developing a more flexible scheme\footnote{Since finishing this paper, we have developed  a more robust  numerical methods which regulates the integral at early times without using a  damping  factor  \cite{paper2}. The $\beta$ regularization is prone to suppressing the 3-point function since pushing $\tau_{early}$ to large negative values make the numerical integrations very time consuming. In practice the results shown here for the step potential may be  $\sim10\%$ lower than their exact values, with a greater suppression at larger values of $k$.}.
As an aside, the imaginary component of the integral goes to
infinity, scaling as $a^2$. However,  as long as we stop the
integration a few efolds after horizon crossing, the integration is
stable.

\section{Results and Discussion} \label{section:results}

\subsection{Shape and Scale}

Every 3-point correlation function has two main attributes: shape and
scale. Unlike previous treatments, we cannot assume that the 3-point
function is scale invariant.  Before we compare the analytical and
numerical results, let us define a more useful parameter than the
rather unwieldy raw 3-point function, 
\begin{equation}
\langle \zeta(\bk_1)\zeta(\bk_2)\zeta(\bk_3)\rangle = (2\pi)^7 N \delta^3(\bk_1+\bk_2+\bk_3)\frac{1}{\Pi_i k_i^3}\CG(\bk_1,\bk_2,\bk_3). \label{eqn:scaling3form}s
\end{equation}
We have factored out the product $1/\Pi_i k_i^3$, since $u(\tau) \sim k^{-3/2}\exp (-ik\tau)$ during
inflation and there are 6 factors of $u$ in the 3-point function. The
function $\CG$ depends on the details of the integrals listed in
Sect. \ref{section:3pts} and $N$ is some constant that we will come
back to later. If the slow roll parameters stay small and constant we
can factor the power spectrum out of $\CG$  \cite{Maldacena:2002vr,Seery:2005wm,Chen:2006nt} 
\begin{equation}
\langle \zeta(\bk_1)\zeta(\bk_2)\zeta(\bk_3)\rangle =
(2\pi)^7\delta^3(\bk_1+\bk_2+\bk_3)(P_k^{\zeta})^2\frac{1}{\Pi_i
k_i^3}\CA(\bk_1,\bk_2,\bk_3)~, \label{eqn:general3ptform}
\end{equation}
where $\CA$ is a quantity of $\CO(k^3)$ and the convention here
follows Ref.~\cite{Chen:2006nt}. In terms of $\CG$, 
\begin{equation}
N\CG = (P_k^{\zeta})^2 \CA.
\end{equation}
Since the power spectrum is
statistically isotropic \cite{Hajian:2003qq}, we assume that this is
also true in the bispectrum and impose rotational
symmetry so that, combined with the conserved momentum constraint, the
3-point correlation function depends only on the amplitudes of the
$k$'s.

Recall the definition of $f_{NL}$, equation (\ref{eqn:fnlansatz}),
which is often used when computing the non-Gaussianities in
the CMB sky
\cite{Komatsu:2002db,Komatsu:2003iq,Spergel:2006hy,Hikage:2006fe}.
Current constraints on $f_{NL}$, from 3 years of WMAP data are  \cite{Creminelli:2006rz}
\begin{equation}
-36<f_{NL}<100. \label{eqn:fnlconstraint}
\end{equation}
As pointed out in \cite{Babich:2004gb}, this ansatz assumes that the 3-point correlation function has the following \emph{local} form
\begin{equation}
\langle \zeta(\bk_1)\zeta(\bk_2)\zeta(\bk_3)\rangle_{local} = (2\pi)^7\delta^3(\bk_1+\bk_2+\bk_3)\left(-\frac{3}{10}f_{NL}(P_k^{\zeta})^2\right)\frac{\Sigma_i k_i^3}{\Pi_i k_i^3}, \label{eqn:localformFNL}
\end{equation}
or, in terms of $\CA$
\begin{equation}
\CA_{local} = -\frac{3}{10}f_{NL}\Sigma_i k_i^3. \label{eqn:ca_local}
\end{equation}
In other words, the constraint (\ref{eqn:fnlconstraint}) is applicable only if the primordial non-Gaussianities have the local form (\ref{eqn:localformFNL}).

Since equation (\ref{eqn:fnlansatz}) does not contain an explicit scale,  
this form is scale-invariant (neglecting the $\CO(\epsilon)$
correction terms)
\begin{equation}
\langle \zeta(\bk_1)\zeta(\bk_2)\zeta(\bk_3)\rangle_{local}  =
\lambda^9 \langle
\zeta(\lambda\bk_1)\zeta(\lambda\bk_2)\zeta(\lambda\bk_3)\rangle_{local}.
\label{eqn:scaleinvariant}
\end{equation}
Note that there is a small violation of  scale-invariance in the 3-point correlation
function stemming from the power spectrum term which we have ignored
in the proceeding argument. In fact, equation (\ref{eqn:localformFNL})
itself hides a further ambiguity: at what value of $k$ are we
evaluating the power spectrum? This   omission is justified if we
assume that the power spectrum is almost scale-invariant, which is
true for most ``standard'' slow roll single scalar field models. However, for
the step potential we are considering, the power spectrum itself undergoes
drastic oscillations of $\CO(1)$. We will come back to this point
later.

\subsection{Quadratic ($m^2\phi^2$) inflation}

We first reproduce known
results for  a simple slow roll inflation model  \cite{Maldacena:2002vr},
\begin{equation}
V(\phi) = \frac{1}{2}m^2\phi^2 \label{eqn:v_sr}
\end{equation}
where $m=10^{-6}M_p$. 

\begin{figure}[ptbh] 
\centerline{\myfigure{4in}{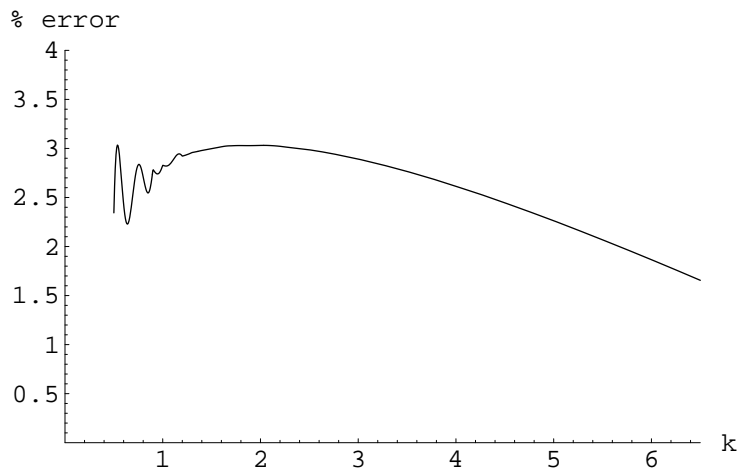}}
\caption{Comparison of $\CA/k^3$ between the analytical slow roll
results of equation (\ref{eqn:ca_sr}) and numerical results from our
code, for the equilateral case where $k_1=k_2=k_3=k$ run from $0.5<k<6.5$ and $\beta = 0.05$. The plot shows
the discrepancy between the two sets of values. The rapid oscillation at small $k$ indicates the breakdown of the current value of $\beta$ to suppress the early time oscillations.
Since the $k$ space spans a few efolds, we have included the contribution from the $\epsilon$ running when computing the analytic estimate but ignored the $\CO(\epsilon^2)$ terms. As we can see, the two results agree to within a few percent.}
\label{fig:errorplot}
\end{figure}

Following \cite{Maldacena:2002vr}, we can write down the explicit form of $\CA$
\begin{equation}
\CA_{SR} = \epsilon \left(-\frac{1}{8}\Sigma_i
k_i^3+\frac{1}{8}\Sigma_{i\neq j}k_i
k_j^2+\frac{1}{k_1+k_2+k_3}\Sigma_{i>j}k_i^2
k_j^2\right)+\eta\left(\frac{1}{8}\Sigma_i k_i^3\right) +
\CO(\epsilon^2) ~, \label{eqn:ca_sr} 
\end{equation}
where $\CO(\epsilon^2)$ terms include $\eta'$.
Comparing this to the local form equation (\ref{eqn:ca_local}) we can
see that for the equilateral case, $-3f_{NL}/10 = 11\epsilon/8+
3\eta/4 = 17\epsilon/8$ where the second equality comes from the fact
that $\eta = 2\epsilon$ for the potential (\ref{eqn:v_sr}).  In the
other extreme of the squeezed triangle case, $-3f_{NL}/10 = 2\epsilon
+\eta = 4\epsilon$. This result was first derived by Maldacena, who
pointed out that for single field slow roll inflationary models,
$f_{NL} = \CO(\epsilon)$ and is thus unobservable. 
We compared our  results to equation (\ref{eqn:ca_sr}), and as Figure \ref{fig:errorplot} shows, our code is accurate to within a few percent, which is within the error expected from the analytical estimate itself.
 
Since the potential has no features, the 3-point correlation
function is essentially ``scale independent''. By this, we mean that triangles which exhibit scaling symmetry posses identical values for their 3-point functions.
 
\subsection{Step potential}

Confident that our code is robust, we now compute the 3-point
correlation functions for the step potential (\ref{eqn:potential}). We
focus on the  best fit model of   Covi {\em et al.\/}
\cite{Covi:2006ci} for a step which affects modes at large angular
scales, namely  where $(c,\phi_s,d) = (0.0018,14.81M_p,0.022)$.   We
choose the same inflaton mass $m = 10^{-6}M_p$ used with the
$m^2\phi^2$ model, so that we can compare the results.

\begin{figure}[tbp] 
\centerline{\myfigure{4in}{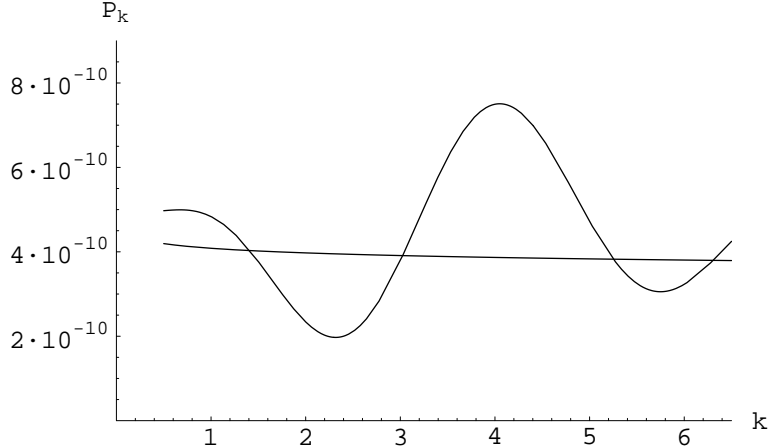}}
\caption{The power spectrum for the standard $m^2\phi^2$ model and the
step potential (\ref{eqn:potential}) for a decade of $k$. In this plot, $m=10^{-6}M_p$ for the $m^2\phi^2$ model while for the step potential we have used the parameters $(c,\phi_s,d) = (0.0018,14.81 M_p,0.022)$.}
\label{fig:plpower}
\end{figure}

Consider the power spectrum, Figure (\ref{fig:plpower}). As first described in
\cite{Adams:2001vc}, the violation of slow roll induces a temporary
growing/decaying behavior in the evolution of the modes before
horizon crossing, resulting in a ``ringing'' of the power
spectrum. Thus we cannot simply factor out the power spectrum from
$\CG$ as we have done in equation (\ref{eqn:general3ptform}); doing so
will introduce spurious contributions to any statistic that we might
obtain.\footnote{For example, dividing it by $\sum_{i>j} P_{k_i}
P_{k_j}$\cite{Lyth:2005fi,Rigopoulos:2005ae}  will entangle 
the ringing in the power spectrum with the ringing in the
bispectrum. The 3-pt correlation function is, in principle, a completely independent statistic from the 2-pt correlation function. In our definition, we make it clear that the only content of the division by $\tilde{P}$ is simply to state that the 3-pt amplitude is roughly the product of 2-pt amplitude. In the future, we envisage that the 3-pt will be part of the set of cosmological parameter space that we can constrain from observations of the CMB and large scale structure, and this definition completely disentangle the 2-pt from the 3-pt and thus avoid any unphysical cross-correlations between the two statistics.}
On the other hand, we expect the 3-point correlation to be of
order $(P_{k}^{\zeta})^2$, so if we set $N=\tilde{P}_k^2$ in equation
(\ref{eqn:scaling3form}) where $\tilde{P}_k \equiv 4\times 10^{-10}$
(i.e. the observed value of the power spectrum), then the
dimensionless quantity 
\begin{equation}
\frac{\CG(k_1,k_2,k_3)}{k_1 k_2 k_3} =
\frac{1}{\delta^3(\bk_1+\bk_2+\bk_3)}\frac{(k_1 k_2
k_3)^2}{\tilde{P}^2(2\pi)^7} \langle
\zeta(\bk_1)\zeta(\bk_2)\zeta(\bk_3)\rangle  \label{eqn:g}
\end{equation}
is what we want to plot. This quantity has the great advantage that
the strong $k^9$ running from the scaling of the 3-point correlation
function (see for example equation (\ref{eqn:scaleinvariant})) is
factored out, so we are left with the scaling that is generated by from the non-linear evolution of the modes.

\begin{figure}[t] 
\myfigure{4in}{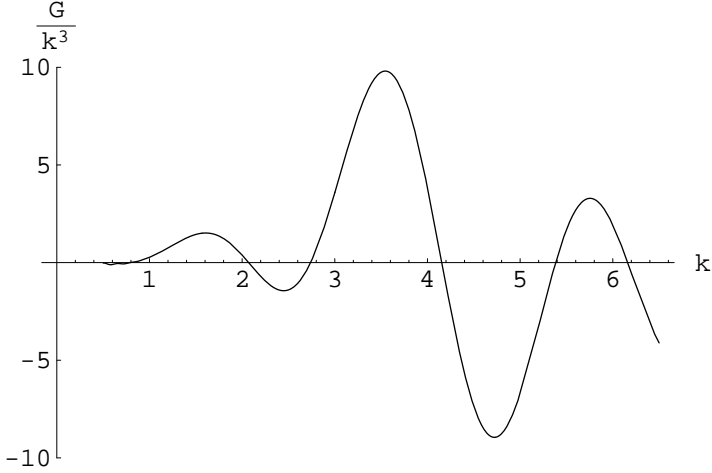}
\caption{The running of non-Gaussianity $\CG/k_1 k_2 k_3$ (in the
equilateral case $k_1=k_2=k_3\equiv k$) 
for the large scale step potential model of
$(c,\phi_s,d)=(0.0018,14.81M_p,0.022)$. For comparison, the standard
slow roll model will yield $\CG/k_1 k_2 k_3 \approx \CO(\epsilon)$.}
\label{fig:plequil3ptlarge}
\end{figure}

Consider the equilateral case, Figure
(\ref{fig:plequil3ptlarge}). Analogously to the power spectrum, the
presence of the step induces a ringing of the equilateral
bispectrum. The ringing frequency in the 3-point function is roughly
1.5 times that in power spectrum, due to the presence of three factors of $\zeta$ in the integrand instead of two.

\begin{figure}[ptbh]
\myfigure{2in}{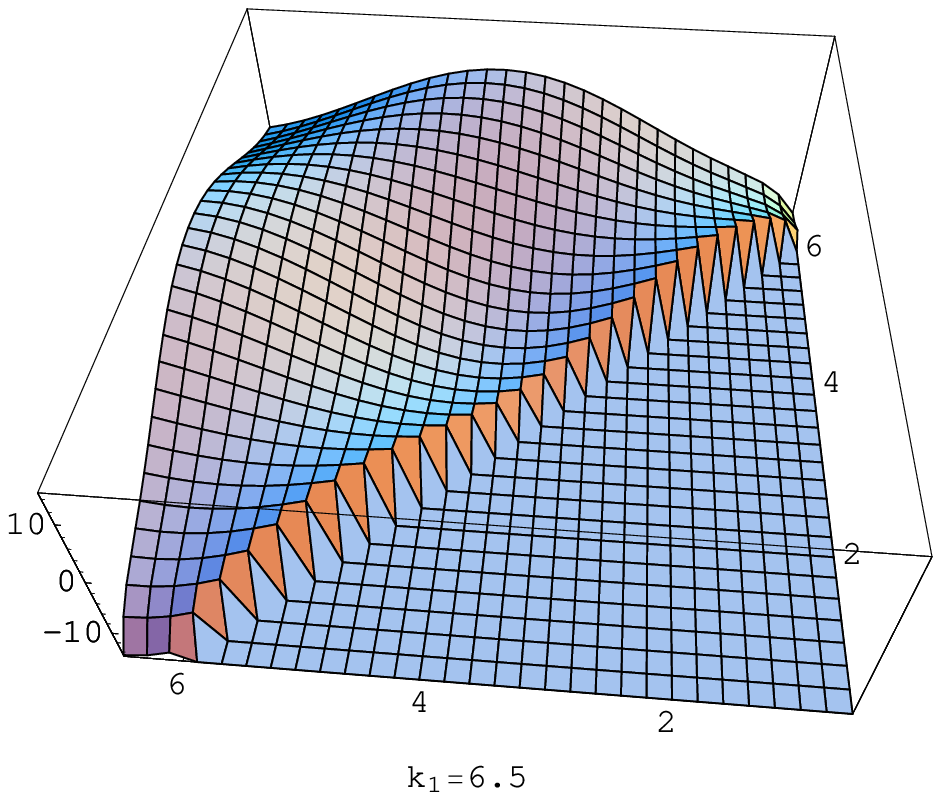}
\myfigure{2in}{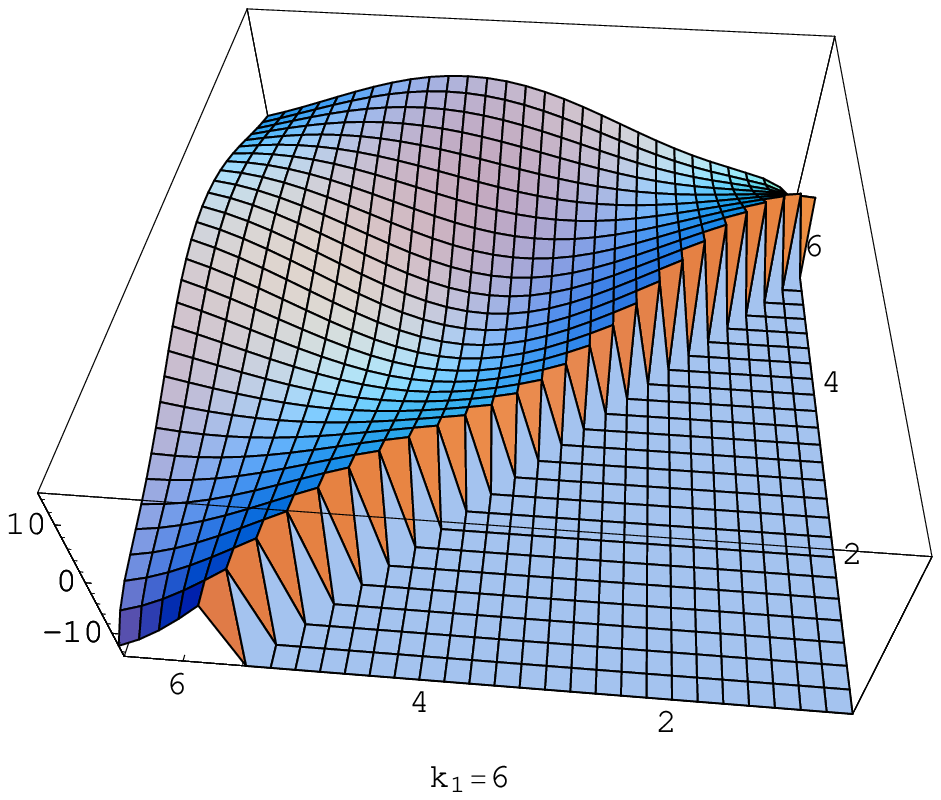}
\myfigure{2in}{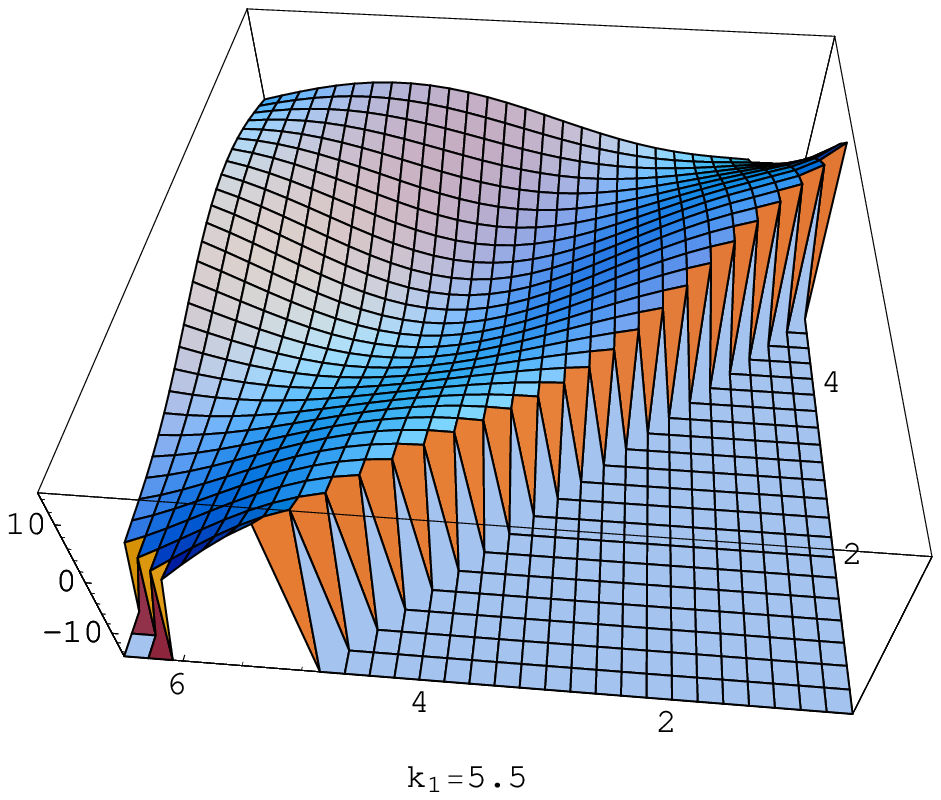}
\myfigure{2in}{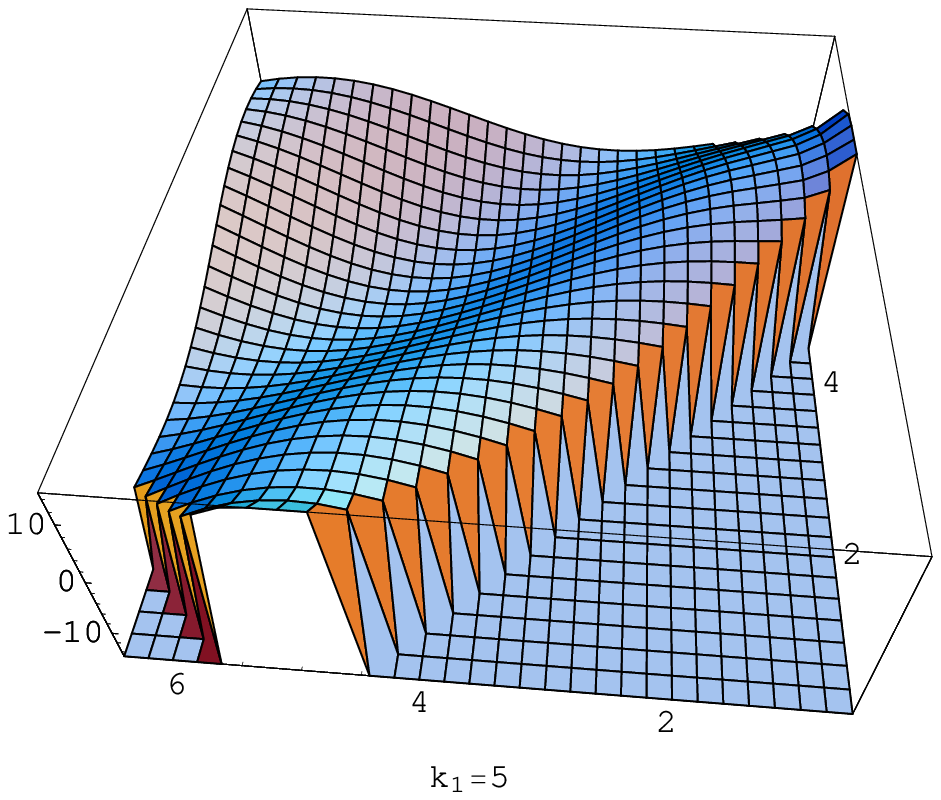}
\myfigure{2in}{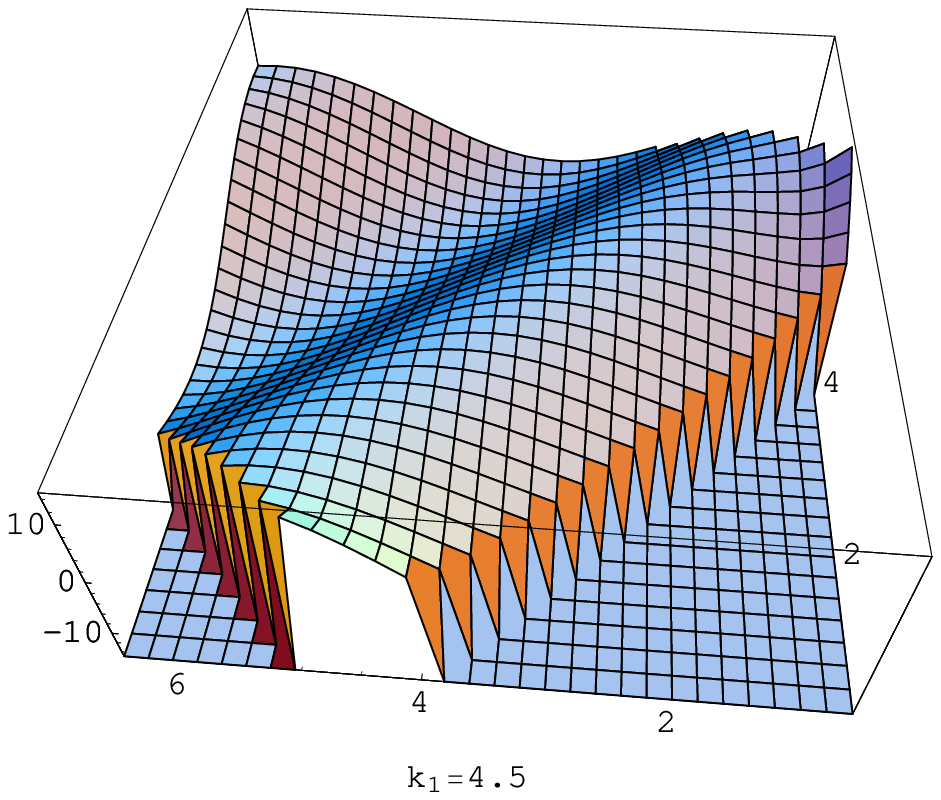}
\myfigure{2in}{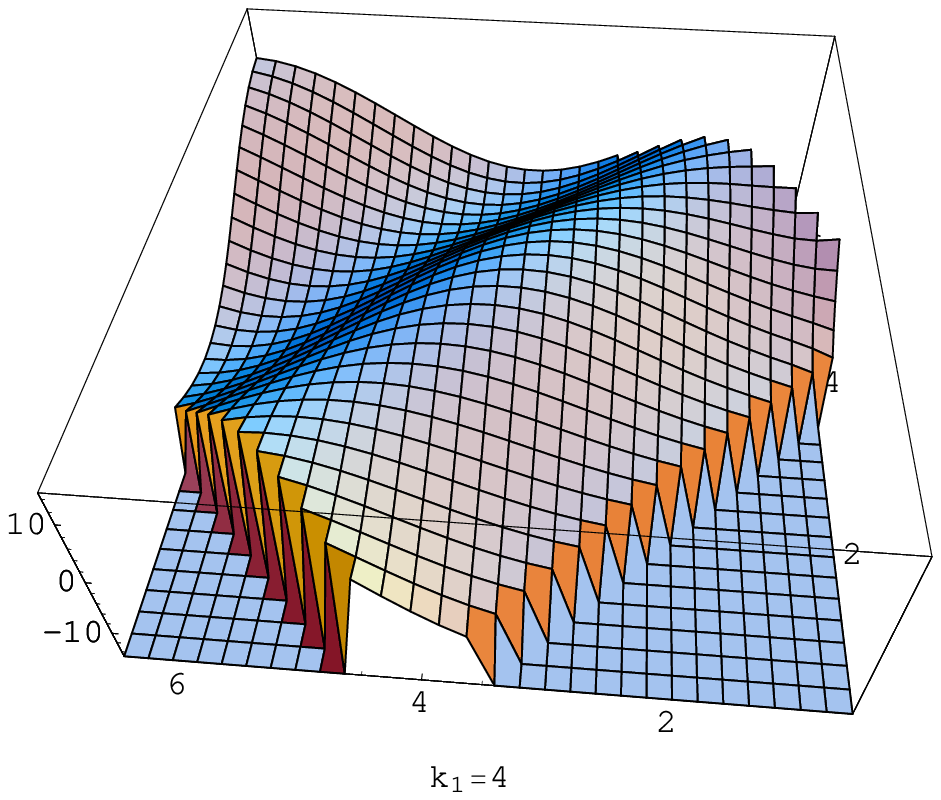}
\myfigure{2in}{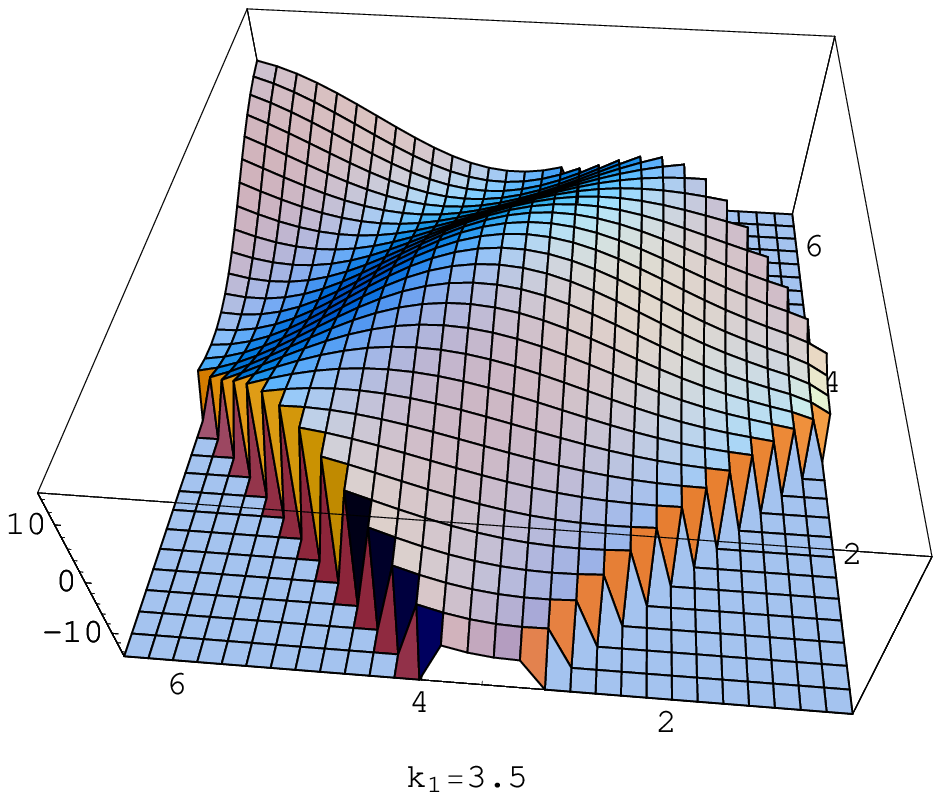}
\myfigure{2in}{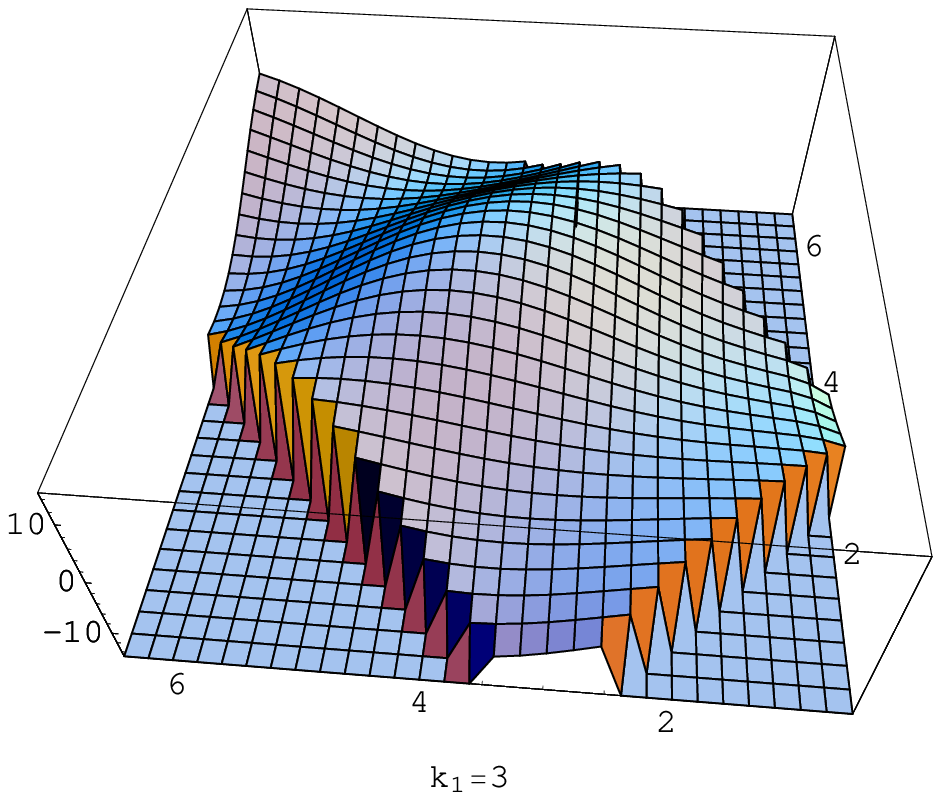}
\myfigure{2in}{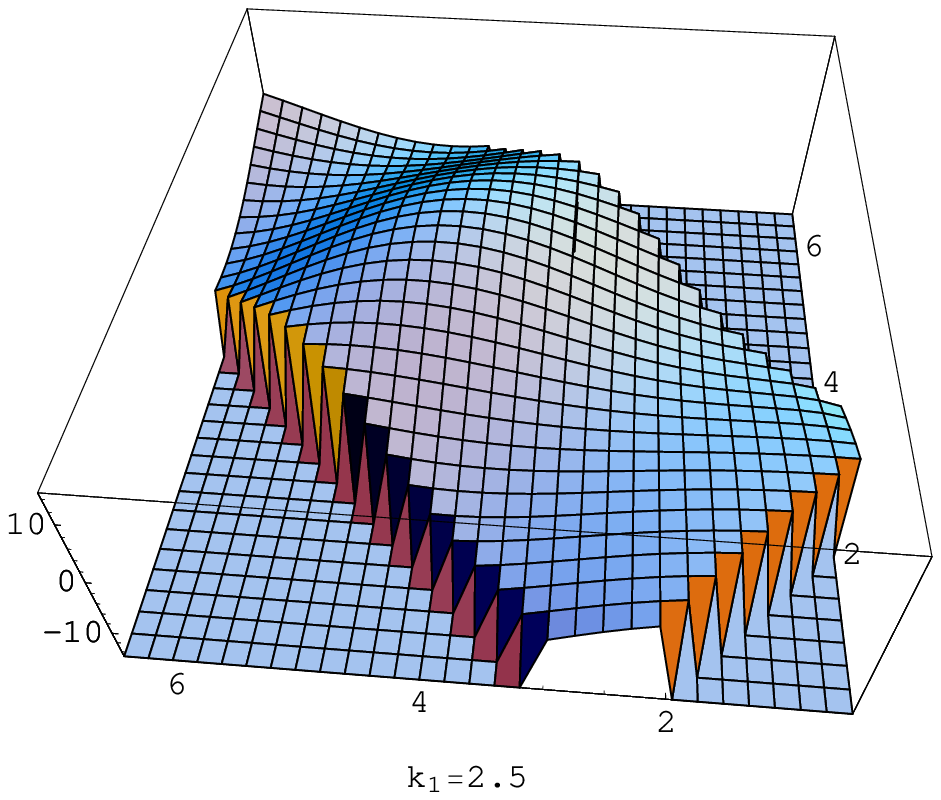}
\caption{The shape of non-Gaussianities 
$\CG/k_1 k_2 k_3$ for the large scale step potential model of
$(c,\phi_s,d)=(0.0018,14.81M_p,0.022)$, with $k_1$ ranging from $6.5$
to $2.5$ going from top left to bottom right. We have set the forbidden triangle regions outside of
$k_i\leq k_j+k_l~,~i\neq j,l$ to $-10$ for visualization purposes. }
\label{fig:large3pt}
\end{figure}

\begin{figure}[ptbh]
\myfigure{3in}{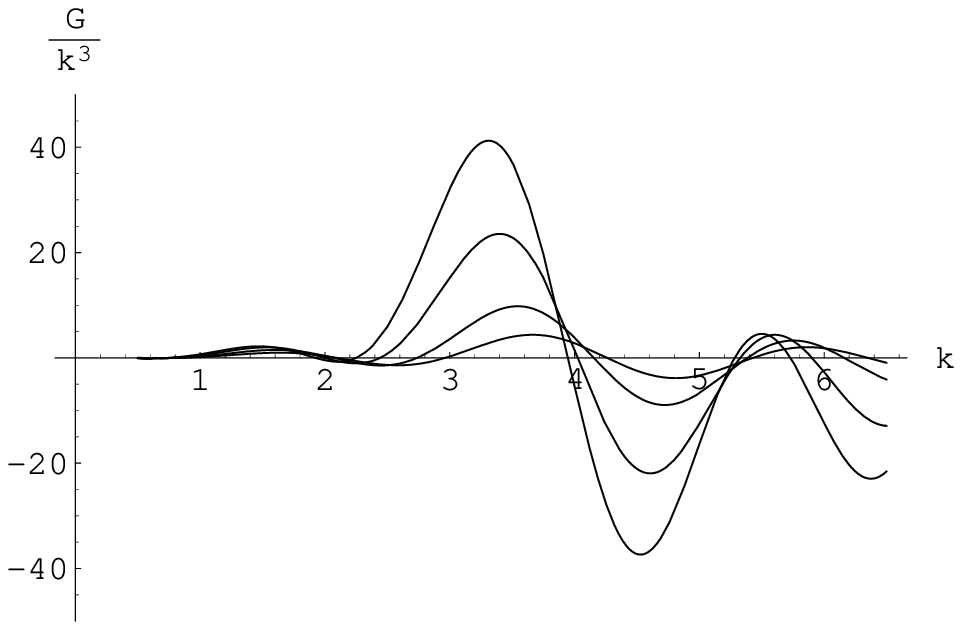}
\myfigure{3in}{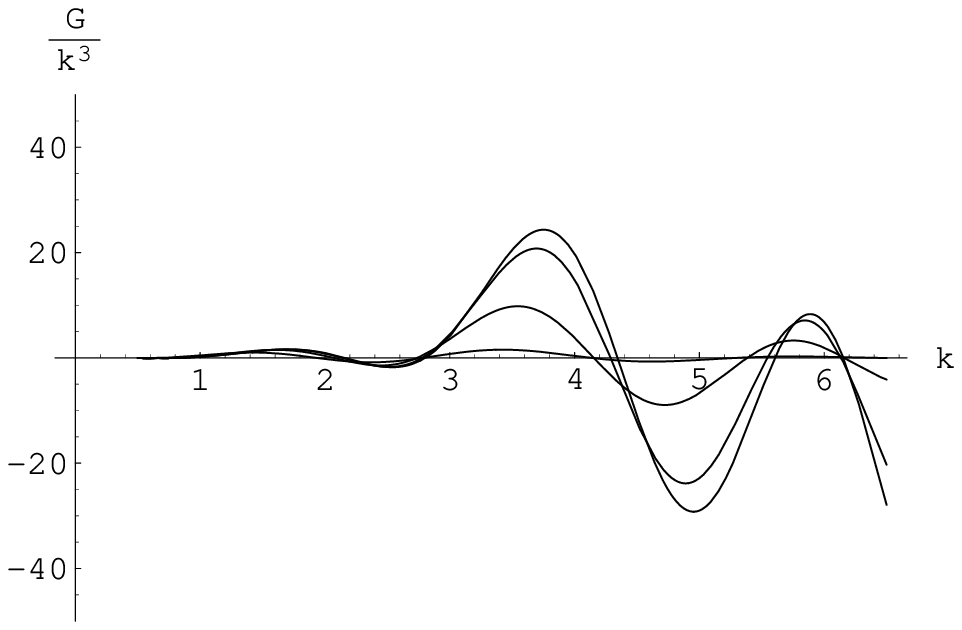}
\caption{Variation of $\CG/k^3$ in response to changing the parameters of the step for the equilateral triangle case. The left (right) plot describes the variation of $c$ ($d$) from the smallest to biggest amplitude in the order $c=0.0009,0.0018,0.0036,0.0054$ fixing $d=0.022$ ($d=0.044,0.022,0.011,0.0074$ fixing $c=0.0018$). We keep the location of the step fixed at $\phi_s = 14.81M_p$.}
\label{fig:cdscaling}
\end{figure}

The plots in Figures \ref{fig:large3pt} illustrate the
complicated landscape of the non-Gaussianities. Compared to the
$\CO(\epsilon)$ amplitude of the $m^2\phi^2$ model, the
non-Gaussianities are magnified several hundredfold and can be larger for other viable step parameters  (Figure \ref{fig:cdscaling}), since we are taking the mid-point values from Covi {\em et al.\/}. As we explained
in Sec.~\ref{section:3pts}, the dominant contribution to the
non-Gaussianities  comes from the $\eta'$ term (Figure
\ref{fig:etadotplot}). 
As we estimated, $\CG/k_1k_2k_3$ 
is of order $7c^{3/2}/(d\epsilon)$ since the contribution from the
$I_{\epsilon^2}$ term is $\CO(\epsilon)$. This is confirmed by 
our numerical results; the shape of the non-gaussianity remains constant but its amplitude is rescaled as we vary $c$ and $d$ together (Figure (\ref{fig:cdscaling})).  

\begin{figure}[ptbh]
\myfigure{2in}{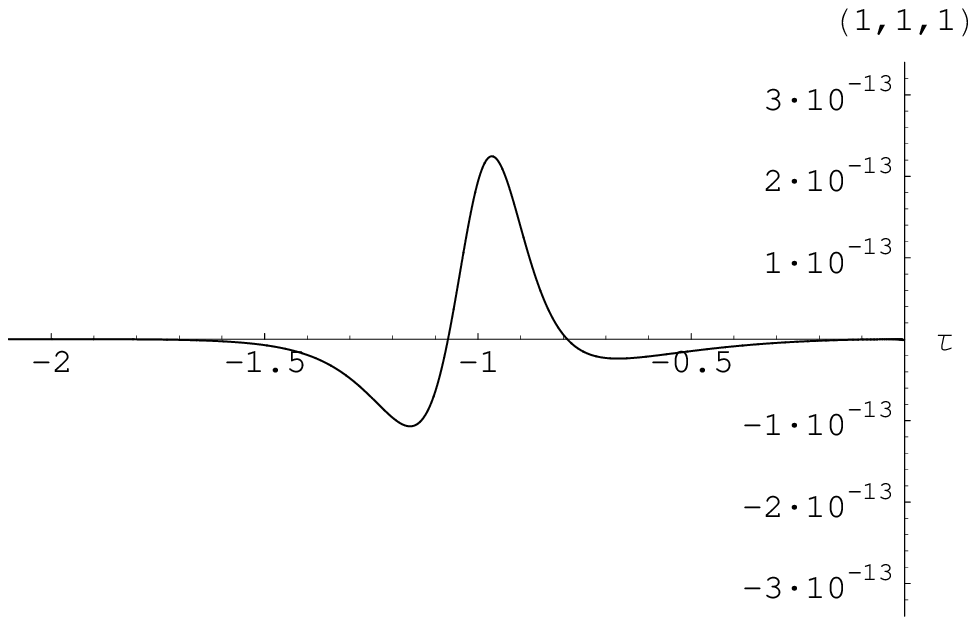}
\myfigure{2in}{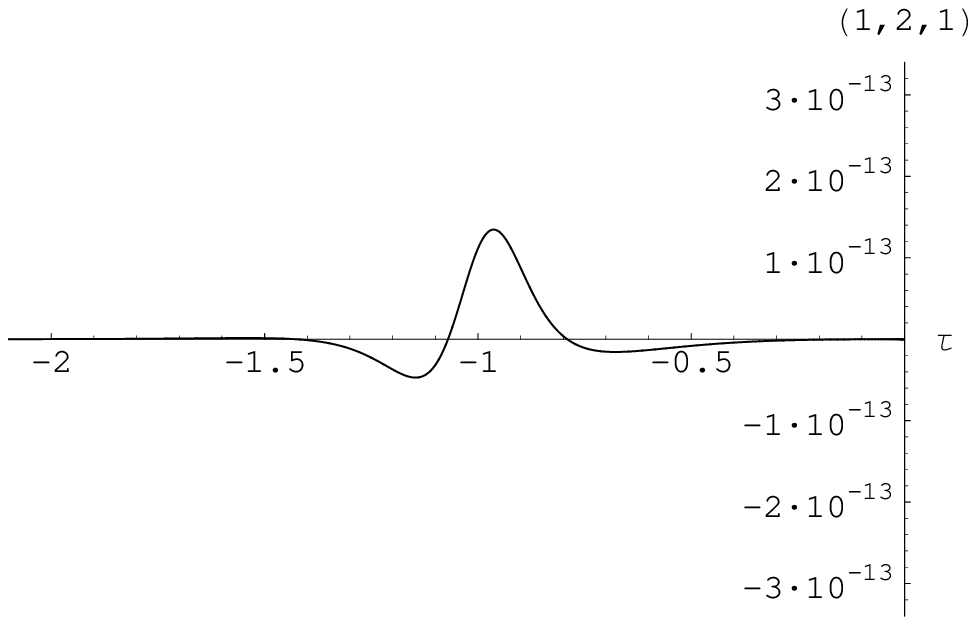}
\myfigure{2in}{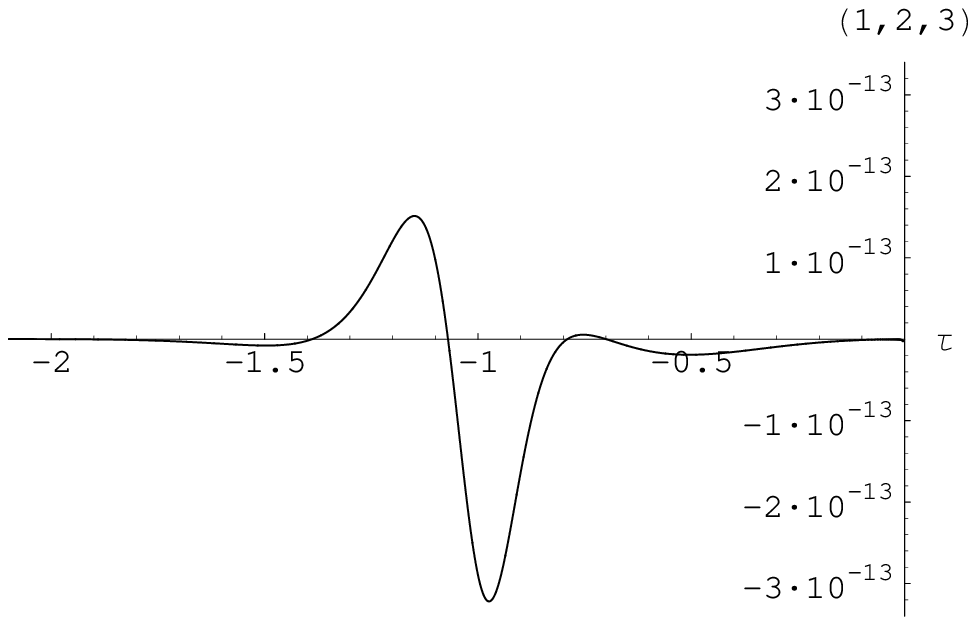}
\myfigure{2in}{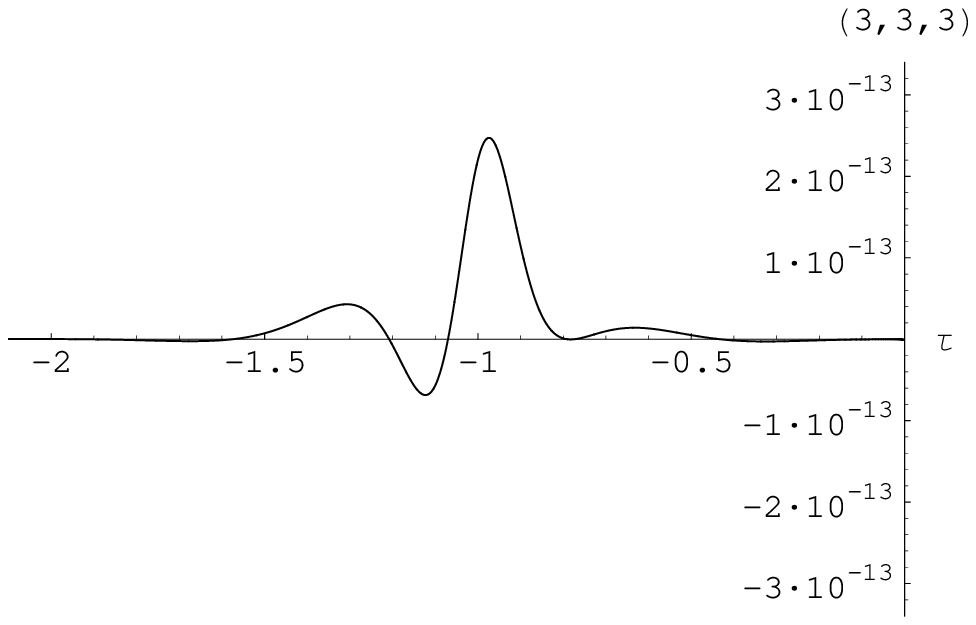}
\myfigure{2in}{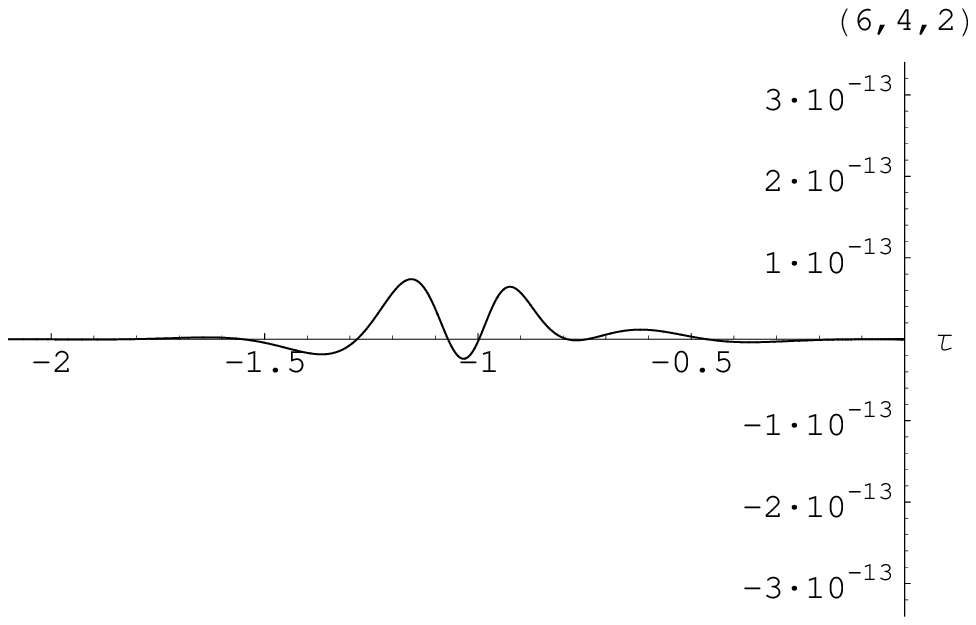}
\myfigure{2in}{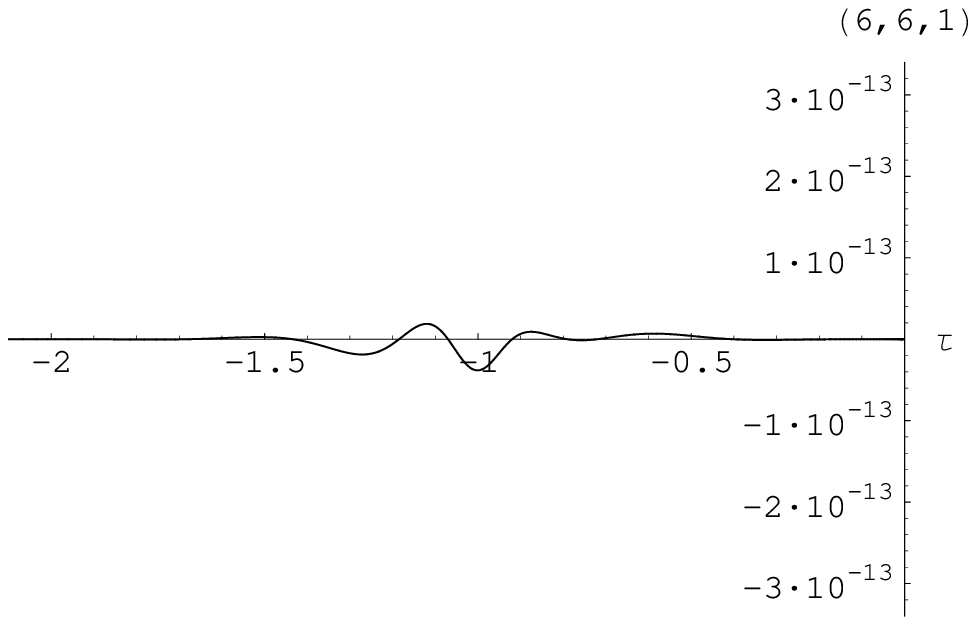}
\caption{The real component of the integrand $(k_1 k_2 k_3)^2 a^2\epsilon \eta' u_1(0)u_2(0)u_3(0) u^*_1u^*_2u^*_3{}'$ plotted
for the modes $(k_1,k_2,k_3)$, with the step occurring around $\tau=-1$
and inflation ending at $\tau=0$. Note that the early time oscillations have been suppressed by the inclusion of a damping factor $\beta$ in the integrand.}
\label{fig:integrals}
\end{figure}

In Figure \ref{fig:integrals}, we plot the integrand 
$(k_1 k_2 k_3)^2 I_{\eta' \epsilon}$ with respect to $\tau$ for the modes which
cross the Hubble horizon around the step. We have suppressed the
early time oscillations via our ``damping trick"  in
order to isolate the actual contribution to the integral. The step
also occurs around this time. Comparing this with the $\eta'$ plot
Figure (\ref{fig:etadotplot}), we see that the oscillatory nature of $
u^*_1u^*_2u^*_3{}'$ modulates the single hump of $\eta'$, yielding
complicated structures. Depending on the combined phases of the modes,
the modulation can be both constructive and destructive, leading to  the ``ringing'' of the primordial bispectrum.
The non-linear interactions between the modes have become dominant due
to the large coupling term $\epsilon \eta'$ around crossing, mixing up
the perturbations in non-trivial manner and generating large
departures from Gaussianity.  This is in contrast  to the ringing of the power
spectrum which is caused by the change in the effective mass of the
modes as they cross the horizon. 

 To bound this specific non-Gaussian signal with current WMAP data, we would have to redo the full sky analysis
using the data for our complicated scale and shape dependent
predictions of the bispectrum. This is not a trivial task and requires
new methods to be developed before we can approach the problem.  A brute force computation of the likelihood is computationally daunting since the
bispectrum here is clearly unfactorizable into a product of separate
integrals over the $k$'s\footnote{A recent paper \cite{Fergusson:2006pr}  has begun to address this important question.}.  
However, we can make some guesses. We have previously alluded to the fact that the statistic (\ref{eqn:g}) is similar to $f_{NL}$. By comparing equation (\ref{eqn:ca_local}) to equation (\ref{eqn:g}), we see that
\begin{equation}
f_{NL} \sim -\frac{10 k_1 k_2 k_3}{3\Sigma_i k_i^3}
\frac{\CG(k_1,k_2,k_3)}{k_1 k_2 k_3}.
\end{equation}
Since the factor $10 k_1 k_2 k_3/3\Sigma_i k_i^3$ is roughly of $\CO(1)$
then $f_{NL} \sim \CG(k_1,k_2,k_3)/k_1 k_2 k_3  \sim \CO(10)$ which is
within sights of the next generation CMB experiments
\cite{Hikage:2006fe}.   However, in our case only a subset of all possible triangles we could draw on the sky have a large 3-point function, which increases the risk that cosmic variance will swamp any signal.  On the other hand, we have a specific template for the 3-point function, which can then be cross-correlated with the 2-point function associated with any putative step. 

\section{Conclusion} \label{section:conclusion}

While  simple models of slow roll inflation do not exhibit observable
non-Gaussianities, the space of models of inflation that can produce a
power spectrum which fits the CMB data is infinite. To discriminate
between these degenerate models, some other independent statistic derived from 
the CMB must be used and the bispectrum is becoming the most promising
candidate for this role.   Unlike the power spectrum, the computation of the bispectrum  for a given 
inflationary model is a messy affair. In this paper, we have developed a
robust numerical code which can compute the 3-point correlation
function for a non-slow roll single field model.  

Using this method, we numerically integrated the 3-point correlation
functions for a class of inflationary models where slow roll is
violated for a brief moment. We show that the addition of this step breaks
the scale invariance of the bispectrum in a non-trivial way, leading
to large non-Gaussianities. This numerical technique can be extended
to solve more complicated problems such as multifield models of
inflation \cite{Battefeld:2006sz,Vernizzi:2006ve} and to compute
higher statistics such as the trispectrum \cite{Seery:2006vu}.

Mirroring the behavior of  the power spectrum, a temporary violation of slow roll
introduces a ringing into the bispectrum, resulting in a structure
much like the transient vibrations of a rug being beaten. These
large non-Gaussianities are generated when the non-linear coupling
between the modes, $\epsilon \eta'$, becomes temporarily large during
this period of non-slow roll behavior. In order to check whether we
can constrain such models with data, a full sky analysis must be
undertaken with either current or future data, a task which is beyond the scope
of this paper. However, we argue that at least for the models
considered in this paper, the non-Gaussianities may be  large
enough to be within reach of the Planck satellite. 

In the near future we can expect even better quality CMB data
to become available, opening up the possibility of cross-correlating higher order statistics such as the bispectrum with the power spectrum. This should provide a powerful
consistency check on any models of inflation with a sharp feature in the potential, even if this  improves the fit to the 2-point
function. Consequently, we have laid the groundwork
for an analog of the ``consistency relationship'' of simple models of slow roll inflation, where the amplitude
and slope of the scalar and tensor spectra are expressed as functions
of just three independent parameters - the observed amplitude of the scalar spectrum, and the first two slow roll parameters, $\epsilon$ and $\eta$.   In this case considered here,  the parameters
that describe any step or other near-discontinuity would fix both
the 2- and 3-point functions, yielding a consistency test for a very broad class of inflationary
models. In practice we could fit separately to the 2- and 3-point functions and then check that the results matched. Alternatively, one could simultaneously include information from both the power spectrum and bispectrum into the cosmological parameter estimation process. In this case, we could fit  directly to the parameters $c$, $d$ and $\phi_s$ (along with the ``average'' slow roll parameters), generalizing the approach of \cite{Peiris:2006sj,Peiris:2006ug}.  Finally, if it ever becomes possible to determine the 3-point function from observations of large scale structure, this information would further tighten the constraints on models where slow roll is briefly violated.

\section{Acknowledgments}

We thank Wayne Hu, Hiranya Peiris, Kendrick Smith, Henry Tye and an anonymous referee for a number of 
useful discussions.  XC thanks the physics department of Yale
university and KITP for their hospitality. 
XC is supported in part by the National Science
Foundation under grant PHY-0355005. 
RE  and EAL are supported in part by  the United States Department of Energy,
grant DE-FG02-92ER-40704. 

\appendix
\section{Other terms}

In Section \ref{section:3pts}, we have given the expressions for the
leading non-Gaussianity in the case when $\eta$ is large but
$\epsilon$ remains small. In this appendix, we give the expressions
for the subleading terms for this case
from the Hamiltonian (\ref{Hint3}).

The subleading contributions come from the first line of (\ref{Hint3}). Contribution from $a^3\epsilon^2 \zeta \dot \zeta^2$ term:
\begin{equation}
2i\int_{-\infty}^{\tau_{end}} d\tau~ a^2 \epsilon^2
\left(\prod_i u_{i}(\tau_{end}) \right) 
\left( u^*_{1} \frac{du^*_{2}}{d\tau} \frac{du^*_{3}}{d\tau} +
{\rm two~perm} \right)
(2\pi)^3 \delta^3(\sum_i \bk_i) + {\rm c.c.} ~.
\label{term2}
\end{equation}
Contribution from $a \epsilon^2 \zeta (\partial \zeta)^2$ term:
\begin{equation}
-2i \int_{-\infty}^{\tau_{end}} d\tau~ a^2 \epsilon^2
\left(\prod_i u_{i}(\tau_{end}) u^*_{i}(\tau) \right) 
\left(\bk_1\cdot \bk_2 + {\rm two~perm} \right)
(2\pi)^3 \delta^3(\sum_i \bk_i) + {\rm c.c.} ~.
\label{term3}
\end{equation}
Contribution from $-2a\epsilon \dot\zeta (\partial\zeta)
(\partial\chi)$ term:
\begin{equation}
-2i\int_{-\infty}^{\tau_{end}} d\tau~ a^2 \epsilon^2
\left(\prod_i u_{i}(\tau_{end}) \right) 
\left( u^*_{1} \frac{du^*_{2}}{d\tau} \frac{du^*_{3}}{d\tau}
\frac{\bk_1\cdot \bk_2}{k_2^2} + {\rm five~perm} \right)
(2\pi)^3 \delta^3(\sum_i \bk_i) + {\rm c.c.} ~.
\nonumber \\
\label{term4}
\end{equation}
The redefinition (\ref{redefinition}) contributes
\begin{equation}
\frac{\eta}{2} ~ |u_{2}|^2 |u_{3}|^2 \Bigg|_{\tau \to \tau_{end}}
(2\pi)^3 \delta^3(\sum_i \bk_i) + {\rm two~perm} ~.
\label{redeterm}
\end{equation}
Note that in Ref.~\cite{Maldacena:2002vr,Seery:2005wm,Chen:2006nt},
where sharp features are absent, these four terms give leading
contributions to non-Gaussianities.

The last two terms in (\ref{Hint3}) are further suppressed by
$\epsilon$, but for completeness we list their contribution in the
following. Contribution from $(\epsilon/2a) \partial\zeta
\partial \chi \partial^2\chi$ term:
\begin{equation}
\frac{i}{2} \int_{-\infty}^{\tau_{end}} d\tau~ a^2 \epsilon^3
\left(\prod_i u_{i}(\tau_{end}) \right) 
\left( u^*_{1} \frac{du^*_{2}}{d\tau} \frac{du^*_{3}}{d\tau}
\frac{\bk_1\cdot \bk_2}{k_2^2} + {\rm five~perm} \right)
(2\pi)^3 \delta^3(\sum_i \bk_i) + {\rm c.c.} ~.
\nonumber \\
\end{equation}
Contribution from $(\epsilon/4a) (\partial^2\zeta)
(\partial\chi)^2$ term:
\begin{equation}
\frac{i}{2} \int_{-\infty}^{\tau_{end}} d\tau~ a^2 \epsilon^3
\left(\prod_i u_{i}(\tau_{end}) \right) 
\left( u^*_{1} \frac{du^*_{2}}{d\tau} \frac{du^*_{3}}{d\tau}
k_1^2 \frac{\bk_2 \cdot \bk_3}{k_2^2 k_3^2} + {\rm two~perm} \right)
(2\pi)^3 \delta^3(\sum_i \bk_i) + {\rm c.c.} ~.
\nonumber \\
\end{equation}

\end{document}